\documentclass{aa}

\usepackage{graphicx}
\usepackage{txfonts}
\begin{document}

\title{Apocenter pile-up and arcs: a narrow dust ring around HD\,129590\thanks{Based on observations made with ESO Telescopes at the Paranal Observatory under programs ID 105.20GP.001 and 109.237K.001.}}
%

\author{J. Olofsson\inst{\ref{inst:MPIA},\ref{inst:NPF},\ref{inst:IFA}}
        \and
        P. Th\'ebault\inst{\ref{inst:LESIA}}
        \and
        A. Bayo\inst{\ref{inst:ESO},\ref{inst:NPF},\ref{inst:IFA}}
        \and
        J. Milli\inst{\ref{inst:IPAG}}
        \and
        R. G. van Holstein\inst{\ref{inst:ESOChile}}
        \and
        Th. Henning\inst{\ref{inst:MPIA}}
        \and
        B. Medina-Olea\inst{\ref{inst:NPF},\ref{inst:IFA}}
        \and
        N. Godoy\inst{\ref{inst:NPF},\ref{inst:IFA}}
        \and
        K. Mauc\'o\inst{\ref{inst:ESO}}
      }
\institute{
    Max Planck Institut f\"ur Astronomie, K\"onigstuhl 17, 69117 Heidelberg, Germany\\\email{olofsson@mpia.de}\label{inst:MPIA}
\and
N\'ucleo Milenio Formaci\'on Planetaria - NPF, Universidad de Valpara\'iso, Av. Gran Breta\~na 1111, Valpara\'iso, Chile\label{inst:NPF}
\and
Instituto de F\'isica y Astronom\'ia, Facultad de Ciencias, Universidad de Valpara\'iso, Av. Gran Breta\~na 1111, Playa Ancha, Valpara\'iso, Chile\label{inst:IFA}
\and
LESIA-Observatoire de Paris, UPMC Univ. Paris 06, Univ. Paris-Diderot, France\label{inst:LESIA}
\and
European Southern Observatory, Karl-Schwarzschild-Strasse 2, 85748 Garching bei M\"unchen, Germany\label{inst:ESO}
\and
Univ. Grenoble Alpes, CNRS, IPAG, 38000, Grenoble, France\label{inst:IPAG}
\and
European Southern Observatory, Alonso de Cordova 3107, Casilla 19001, Vitacura, Santiago, Chile\label{inst:ESOChile}
}


\abstract{Observations of debris disks have significantly improved over the past decades, both in terms of sensitivity and spatial resolution. At near-infrared wavelengths, new observing strategies and post-processing algorithms allow us to drastically improve the final images, revealing faint structures in the disks. These structures inform us about the properties and spatial distribution of the small dust particles.}
{We present new $H$-band observations of the disk around the solar type star HD\,129590, which display an intriguing arc-like structure in total intensity but not in polarimetry, and propose an explanation for the origin of this arc.}
{Assuming geometric parameters for the birth ring of planetesimals, our model provides the positions of millions of particles of different sizes to compute scattered light images. The code can either produce images over the full size distribution or over several smaller intervals of grain sizes.}
{We demonstrate that if the grain size distribution is truncated or strongly peaks at a size larger than the radiation pressure blow-out size we are able to produce an arc quite similar to the one detected in the observations. If the birth ring is radially narrow, given that particles of a given size have similar eccentricities, they will have their apocenters at the same distance from the star. Since this is where the particles will spend most of their time, this results in a ``apocenter pile-up'' that can look like a ring. Due to more efficient forward scattering this arc only appears in total intensity observations and remains undetected in polarimetric data, in good agreement with our observations.}
{This scenario requires sharp variations either in the grain size distribution or for the scattering efficiencies $Q_\mathrm{sca}$ (or a combination of both). Alternative possibilities such as a wavy size distribution and a size-dependent phase function are interesting candidates to strengthen the apocenter pile-up. We also discuss why such arcs are not commonly detected in other systems, which can mainly be explained by the fact that most parent belts are usually broad.}

\keywords{Stars: individual (HD\,129590) -- circumstellar matter -- Techniques: high angular resolution}

\maketitle
%

\section{Introduction}

Since the first detection of circumstellar dust around a main sequence star other than the sun (\citealp{Aumann1984}), observations of debris disks have kept on improving. Starting from unresolved far-infrared (IR) photometry, we now have access to optical and near-IR extreme adaptive optics instruments such as the Spectro-Polarimetric High-contrast Exoplanet REsearch (SPHERE, \citealp{Beuzit2019}), the Gemini Planet Imager (\citealp{Macintohs2014}), or the Subaru Coronagraphic Extreme Adaptive Optics system (\citealp{Jovanoic2015}), as well as (sub-)mm interferometric high angular resolution observations with the Atacama Large Millimeter/submillimeter Array (ALMA). These instruments and facilities offer an angular resolution of a few tens of milli-arcsec (mas), allowing us to spatially resolve debris disks with exquisite details, revealing the birth ring where planetesimals are colliding and releasing the small dust particles that we can observe.

Thanks to the high angular resolution of near-IR scattered light observations, we start to detect faint sub-structures in some of those disks. These structures can range from multiple belts (HD\,131835, \citealp{Feldt2017} and HD\,120326, \citealp{Bonnefoy2017}), spiral arm, clumps, and rings (TWA\,7, \citealp{Olofsson2018}, \citealp{Ren2021}), or large-scale extended swept-back wings or halos (HD\,61005 and HD\,32297, \citealp{Schneider2014} and HR\,4796, \citealp{Schneider2018}). In this paper we investigate what can be learned from the detection of some of these sub-structures, applied to the case of HD\,129590.

HD\,129590 is a G3V star, located at $136.32 \pm 0.44$\,pc (\citealp{Gaia2016,Gaia2020}). It is a member of the Upper Centaurus Lupus Sco-Cen region (\citealp{dZ1999}; \citealp{Chen2011}). It hosts a bright debris disk (fractional luminosity $f_\mathrm{disk} \sim 5 \times 10^{-3}$, \citealp{Jang2015}), which was first spatially resolved at near-IR wavelengths, in total intensity, by \citet{Matthews2017}. The disk is highly inclined ($i \sim 75^{\circ}$) with a radius of $60-70$\,au ($58$ and $67.7$\,au correcting for the previous estimate of the distance of $141$\,pc). \citet{Olofsson2022} presented polarimetric observations of the disk, and found a slightly larger inclination ($i \sim 82^{\circ}$) and smaller radius ($47.6$\,au). From ALMA observations, \citet{Kral2020} also reported the detection of CO gas in the system ($\sim 10^{-4}$\,$M_\oplus$), the first disk around a solar type star to harbor cold gas.

In this paper we present new observations obtained with the SPHERE instrument using the star-hopping method (\citealp{Wahhaj2021}). The observations were processed using recent post-processing algorithms allowing us to retrieve faint structures in the total intensity image. The aim of this paper is not to perform detailed modeling of the observations but rather discuss what we can learn about the properties of the disk from the detection of such structures.

\section{Observations and data reduction}\label{sec:data}

\begin{figure*}
  \centering
  \includegraphics[width=\hsize]{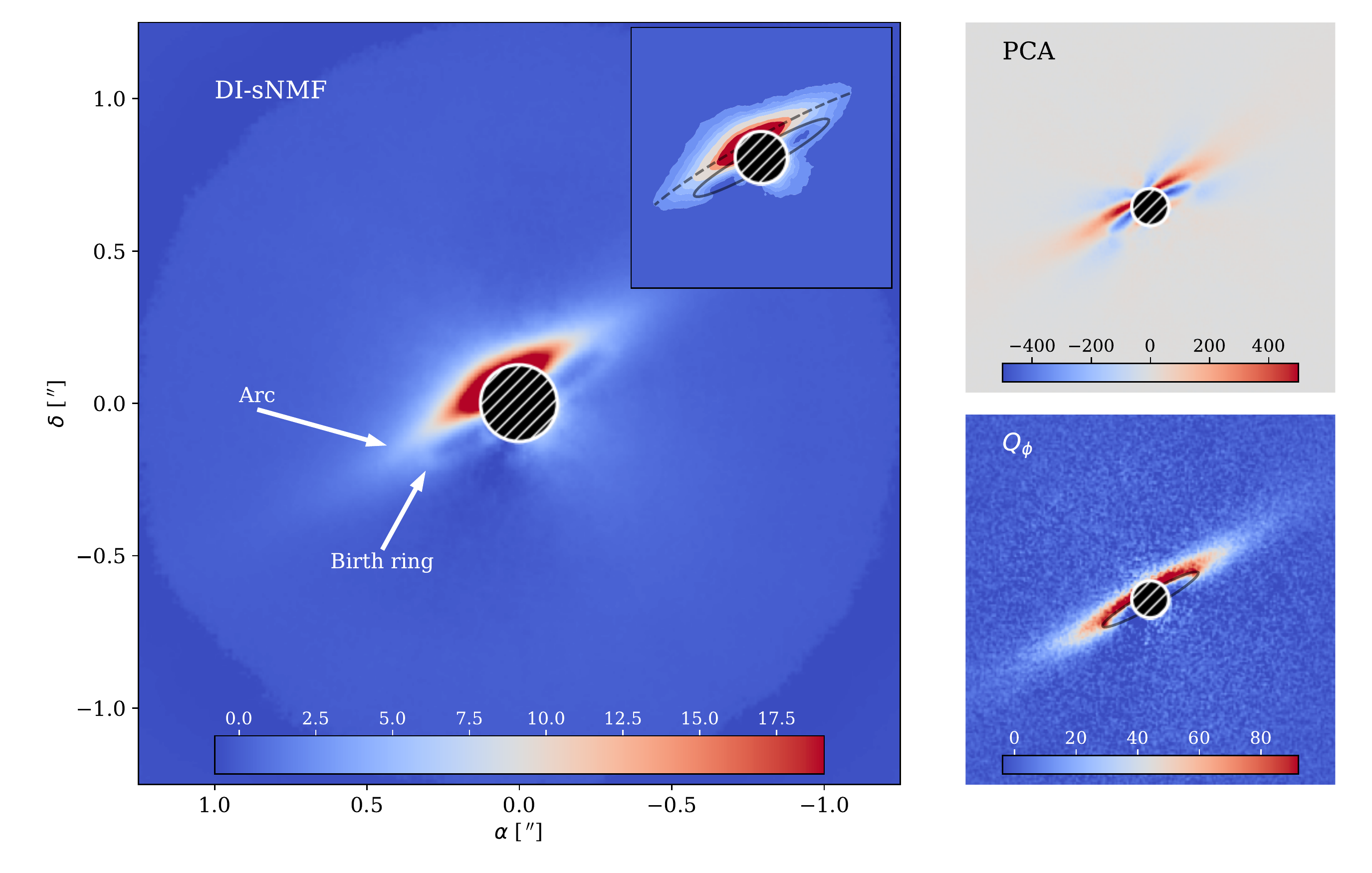}
    \caption{SPHERE-IRDIS observations of the disk around HD\,129590, showing the DI-sNMF (total intensity, left), PCA (total intensity, top right) and $Q_\phi$ (polarized intensity, bottom right) reductions. The detection of the birth ring and the presence of a faint arc are highlighted by two arrows on the left panel. The inset on the left panel also shows the DI-sNMF reduction with the birth ring (best fit from \citealp{Olofsson2022}) and the arc highlighted. The location of the birth ring is also shown on the $Q_\phi$ image as a solid black line. For each image, the central mask has a radius of $0.125$\arcsec, and the color scale is linear. North is up, east is left and the pixel scale is of $12.26$\,mas.}
  \label{fig:data}
\end{figure*}

The dataset used in this study was obtained using the SPHERE IRDIS instrument (\citealp{Dohlen2008}), using the dual-beam polarimetric imaging mode (DPI, \citealp{deBoer2020}, \citealp{vanHolstein2020}) in pupil tracking (\citealp{vanHolstein2017}), in the $H$ band ($\lambda_\mathrm{c} = 1.62$\,$\mu$m). The observations were performed making use of the star-hopping mode of the instrument (\citealp{Wahhaj2021}), where the instrument hops back and forth between the science target and a reference star. The motivation is to build a reference coronagraphic image from the reference calibrator star without any circumstellar material, to be subtracted to the science cube to reveal faint signal in total intensity, in the immediate vicinity of the star.

\subsection{SPHERE total intensity images}

Most routinely, when studying debris disks, the pupil-stabilized observations are processed using angular differential imaging (\citealp{Marois2006}) with a principal component analysis (PCA), however, self-subtraction can significantly alter parts of the disk (especially along the projected semi-minor axis, \citealp{Milli2012}). Self-subtraction can be mitigated using approaches such as forward modeling (e.g., \citealp{Olofsson2016}; \citealp{Chen2020}; \citealp{Mazoyer2020}), nonetheless it remains a model-dependent approach. In the past years, several algorithms became available to process observations without having to rely on a model of the disk, and lowering the impact of self-subtraction, such as \texttt{GreeDS} (\citealp{Pairet2018}), \texttt{REXPACO} (\citealp{Flasseur2021}), \texttt{MAYONNAISE} (\citealp{Pairet2021}), \texttt{mustard} (\citealp{Juillard2022}), or performing data imputation with sequential non-negative matrix factorization (DI-sNMF \citealp{Ren2020}). These post-processing algorithms show promising results and allow us to recover the signal from the disk without suffering from most of the drawbacks of PCA. 

In this study, we processed our total intensity observations using the DI-sNMF algorithm, making use of reference star differential imaging (RDI). In programs 105.20GP.001 and 109.237K.001, we observed four reference stars, HD\,117255, HD\,129280, HD\,158018, and HD\,191131 with the same observing mode as for HD\,129590. After reducing the data using the \texttt{IRDAP}\footnote{Available at \url{https://irdap.readthedocs.io/en/latest/}} package described in \citet[][version 1.3.4]{vanHolstein2020}, we obtained a master calibrator cube of $222$ reference coronagraphic images that can be used to perform RDI. Since the observations have different integration times, each frame was normalized by its corresponding integration time prior to constructing the cube. From this cube, we used the DI-sNMF package\footnote{Available at \url{https://github.com/seawander/nmf_imaging}} to construct the NMF components (a non-orthogonal, non-negative basis built from reference images, see \citealp{Ren2018}). These components are then used on the science data cube ($32$ individual frames), and a numerical mask of $1\arcsec$ in radius, centered on the star, is applied to exclude the disk and perform the data imputation. We used $50$, $75$, $100$, and $200$ components for the reduction, and the differences between the final reductions are minute. We therefore used the reduction with $75$ components. The left and top right panels of Figure\,\ref{fig:data} show the DI-sNMF and PCA\footnote{The PCA was performed on the science cube itself, not making use of the calibrator cube. In Figure\,\ref{fig:data}, we used two components.} reduction respectively, illustrating the significant differences between the two reductions. Not only is there almost no negative signal in the DI-sNMF reduction but the flux of the projected semi-minor axis is much better recovered compared to the PCA reduction where it is completely lost to self-subtraction effects. For the rest of this study, when referring to the total intensity observations, this means the DI-sNMF reduction.

\subsection{Linear polarimetric images}

To obtain the images in linearly polarized light,we used the same reduction as the one presented in \citet{Olofsson2022}, for which we had also used the \texttt{IRDAP} package. The reduction yielded the $Q_\phi$ and $U_\phi$ images, the former containing the signal from the disk and the latter can be used as a proxy for the uncertainties, since the $U_\phi$ image does not contain astrophysical signal (assuming single scattering by particles). The $Q_\phi$ image is shown in the bottom right panel of Figure\,\ref{fig:data}.

\subsection{Detection of an arc in total intensity}

Figure\,\ref{fig:data} shows some interesting differences between the total and polarized intensity observations. The $Q_\phi$ image is rather straightforward to explain. The birth ring, where planetesimals collide with each other, is well resolved, and the disk extends along the major axis of the disk into a halo. This halo traces the small dust particles that are sent on eccentric orbits due to radiation pressure (or a combination of radiation pressure and gas drag, \citealp{Kral2020} and \citealp{Olofsson2022}). In polarimetric observations, forward scattering is not as strong as in total intensity observations. The polarized phase function either peaks (or plateaus) around scattering angles\footnote{The complementary of the angle between the star, the dust particle, and the observer.} in the range $60-90^{\circ}$, or slowly decreases with increasing scattering angles (\citealp{Olofsson2022}). Combined with an increase in column density along the major axis due to the inclination of the disk (\citealp{Olofsson2020}), this renders the major axis of the disk brighter than its minor axis. On the bottom right panel of Fig.\,\ref{fig:data} we also show as a solid black line the location of the birth ring reported in \citet{Olofsson2022}.

On the other hand, the total intensity observations display different structures that are not visible in the $Q_\phi$ image. The birth ring is also detected for a wide range of scattering angles, as we can even detect the back side of the disk (the front side being located towards the north-northeast). But, we can also see an ``arc''-like structure, that seems to be parallel to the direction of the major axis, but slightly offset towards the north, begging the question of its origin. The inset in the left panel of Fig.\,\ref{fig:data} shows contours of the total intensity image and the location of the birth ring, as inferred from the polarimetric data, is shown with a solid black line. We also highlight the location of the arc with a dashed black line. It should be noted that this arc also seems to be detected in the reduction presented in \citet[][see their Fig.\,4]{Matthews2017}, but was not discussed. To assess whether the detection of this arc is an artifact of the DI-sNMF reduction, Figure\,\ref{fig:mayo} shows alternative reductions of the science data cube using other algorithms (\texttt{GreeDS} and \texttt{MAYONNAISE}), in which similar structures are recovered.

\section{A possible explanation: apocenter pile-up}

\subsection{A secondary ring}

One possible explanation for the morphology of the disk as seen in the total intensity observations would be to invoke the presence of an outer disk (as discussed in \citealp{Bonnefoy2017} for HD\,120326, see also Section\,\ref{sec:otherdisks}). Due to projection effects for an inclined system, the outer belt could appear as slightly overlapping the inner one, therefore appearing as an arc. 

To fit the spectral energy distribution, \citet{Chen2014} required two Planck functions at different temperatures, but both \citet{Ballering2013} and \citet{Jang2015} could reproduce it using only one temperature (see discussion in \citealp{Matthews2017}). Furthermore, the polarimetric observations are of good quality and the ``inner'' disk is well detected, while there is no indication for the presence of an outer disk in this data set. Ultimately, we would need high angular resolution observations with ALMA (in \citealp{Kral2020}, the beam size was larger than $1\arcsec$) to confirm that there is a single belt around HD\,129590, but the scenario of two belts seems unlikely. It would require that the dust particles in the outer ring have a near zero degree of polarization, so that they scatter enough stellar light to be detected in total intensity while remaining undetected in polarized light. Given how bright the disk is in the $Q_\phi$ image, this would imply drastically different dust properties for the inner and outer belts. In the following, we will assume that there is only one birth ring around HD\,129590 and try to explore alternative explanations for the presence of the arc-like structure.

\subsection{Radiation pressure and radial size segregation}\label{sec:pfunc}

\begin{figure*}
  \centering
  \includegraphics[width=\hsize]{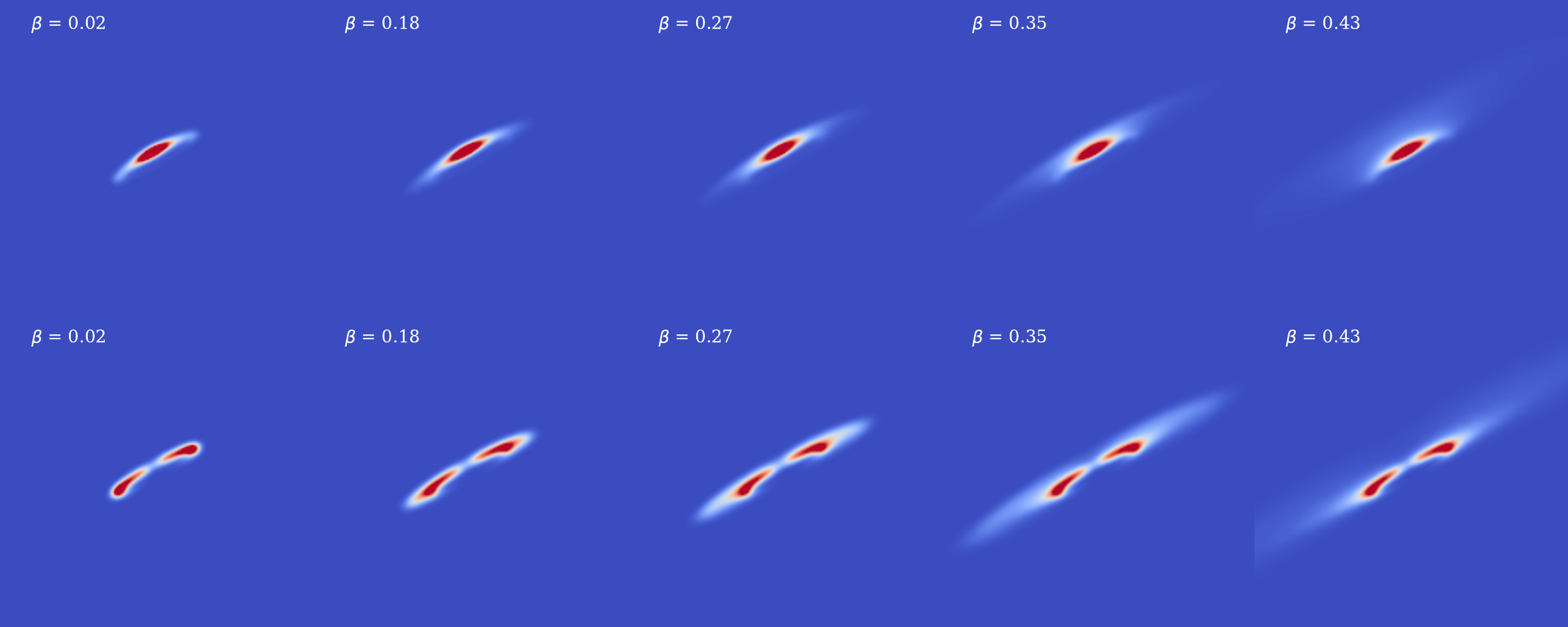}
    \caption{Simulated disk images for total intensity (top panels) and polarized intensity (bottom panels) for different narrow intervals of $\beta$, and the average value of $\beta$ is reported in each panel. The value of the asymmetry parameter $g$ is set to $0.7$ for all images. The scaling is linear and adjusted to the $99.9$ percent for all frames. The images are convolved with a 2D gaussian with a standard deviation of $2$\,pixels.}
  \label{fig:adi_dpi}
\end{figure*}

As previously mentioned, the phase functions in total and polarized intensity are expected to be quite different. Here, we therefore explore if the phase function can explain why we detect an arc-like structure in the total intensity observations while it is not seen in the polarimetric data. We used the same code as the one described in \citet{Olofsson2022b}\footnote{Available at \url{https://github.com/joolof/betadisk}}, which can compute scattered light images for different grain sizes. We refer the interested reader to the original paper for additional details, but in short, the code launches millions of particles that are parametrized by their $\beta$ ratio (the ratio between radiation pressure and gravitational forces, $\propto 1/s$, where $s$ is the particle size). All the particles are initially released from the birth ring, which follows a normal distribution centered at $a_0$ with a standard deviation $\delta_\mathrm{a}$, and the code computes the updated orbital parameters for each particle, including the effect of radiation pressure. Large dust particles (small $\beta$) will remain in the vicinity of the birth ring, while small particles (large $\beta$) will be set on highly eccentric orbits. We only consider grains that remain gravitationally bound to the star, and do not include unbound grains ($\beta \geq 0.5$ if initially released from a circular orbit). The code accounts for the illumination factor ($\propto 1/r^2$), the geometric cross-section of the particles ($\propto 1/\beta^2$ since $\beta \propto 1/s$), the size distribution ($\mathrm{d}n(\beta) \propto \beta^{1.5} \mathrm{d}\beta$, equivalent to $\mathrm{d}n(s) \propto s^{-3.5}\mathrm{d}s$, \citealp{Dohnanyi1969}), and a ``correction factor'' $\alpha$. The last term is necessary because particles set on eccentric orbits will travel in regions of very low density, and therefore their lifetime can be significantly increased (see \citealp{Strubbe2006}, \citealp{Thebault2008}).

The code can also take a phase function as an input, which can follow the Henyey-Greenstein (HG) approximation (\citealp{Henyey1941}). In total intensity, the phase function takes the following form

\begin{equation}\label{eqn:pfunc}
  S_{11}(\theta) = \frac{1}{4 \pi} \frac{1 -g^2}{(1 + g^2 - 2g\cos(\theta))^{3/2}},
\end{equation}

where $g$ is the asymmetry parameter ($-1<g<1$), and $\theta$ the scattering angle. In polarized intensity, the polarized phase function takes the form of 

\begin{equation}\label{eqn:pfuncp}
  S_{12}(\theta) = S_{11}(\theta)\frac{1-\cos^2(\theta)}{1+\cos^2(\theta)}.
\end{equation}

For all the particles, the code draws values for $\beta$ between $\beta_\mathrm{min}$ and $\beta_\mathrm{max}$ (usually $0.01$ and $0.49$), but it is possible to bin the $\beta$ values in different $n_\mathrm{g}$ intervals. Figure\,\ref{fig:adi_dpi} shows different images for $5$ different $\beta$ intervals (with $n_\mathrm{g} = 12$) in total and polarized intensity (top and bottom panels, respectively). The parameters for the birth ring are taken from \citet{Olofsson2022}, with the inclination $i = 82^\circ$, the position angle $\phi = -60.6^\circ$, the semi-major axis $a_0 = 0.35\arcsec$ ($47.7$\,au at a distance of $136.32$\,pc), the width of the birth ring $\delta_\mathrm{a} = 0.025\arcsec$ ($3.4$\,au), and we used a value for $g$ of $0.7$ for both the total and polarized intensity. While \citet{Matthews2017} found a value for the asymmetry parameter $g$ closer to $0.4-0.5$, our choice for $g = 0.7$ is further motivated by the analysis presented in Section\,\ref{sec:size_pfunc}. For comparison, Figure\,\ref{fig:adi_dpi05} shows images similar as those presented in Fig.\,\ref{fig:adi_dpi} for $g = 0.5$. For the vertical structure of the disk, we assumed a normal distribution with a constant aspect ratio and an opening angle of $0.02$\,radians, based on the results of \citet{Olofsson2022} where they found the disk to be vertically thin.

For $\beta = 0.02$, the image traces the birth ring as these large particles do not feel strong radiation pressure. Because of forward scattering, the semi-minor axis of the disk appears bright in total intensity, while the major axis is best traced in polarized light. As $\beta$ increases (up to $\beta \sim 0.35$), we start to see the extended halo becoming more and more prominent in the images as a result of the particles' increased apocenter distances. In polarized intensity, because of the shape of the phase function, this halo extends in the continuity of the major axis of the disk. On top of the phase function itself, as mentioned previously, the major axis is also brighter simply because of projection effects due to the high inclination of the disk. We probe a larger column density along the major axis compared to the minor axis of the disk. However, we can observe that the situation is quite different in total intensity. Due to the stronger forward scattering, the front side of the minor axis becomes brighter than the major axis, and for the images with $\beta = 0.27$ and $0.35$ an arc quite similar to the one seen in the observations becomes noticeable. Interestingly, for $\beta = 0.43$ (and larger), the halo becomes so extended that it appears relatively faint, and we mostly, once again, see the birth ring of the disk.

\begin{figure}
  \centering
  \includegraphics[width=\linewidth]{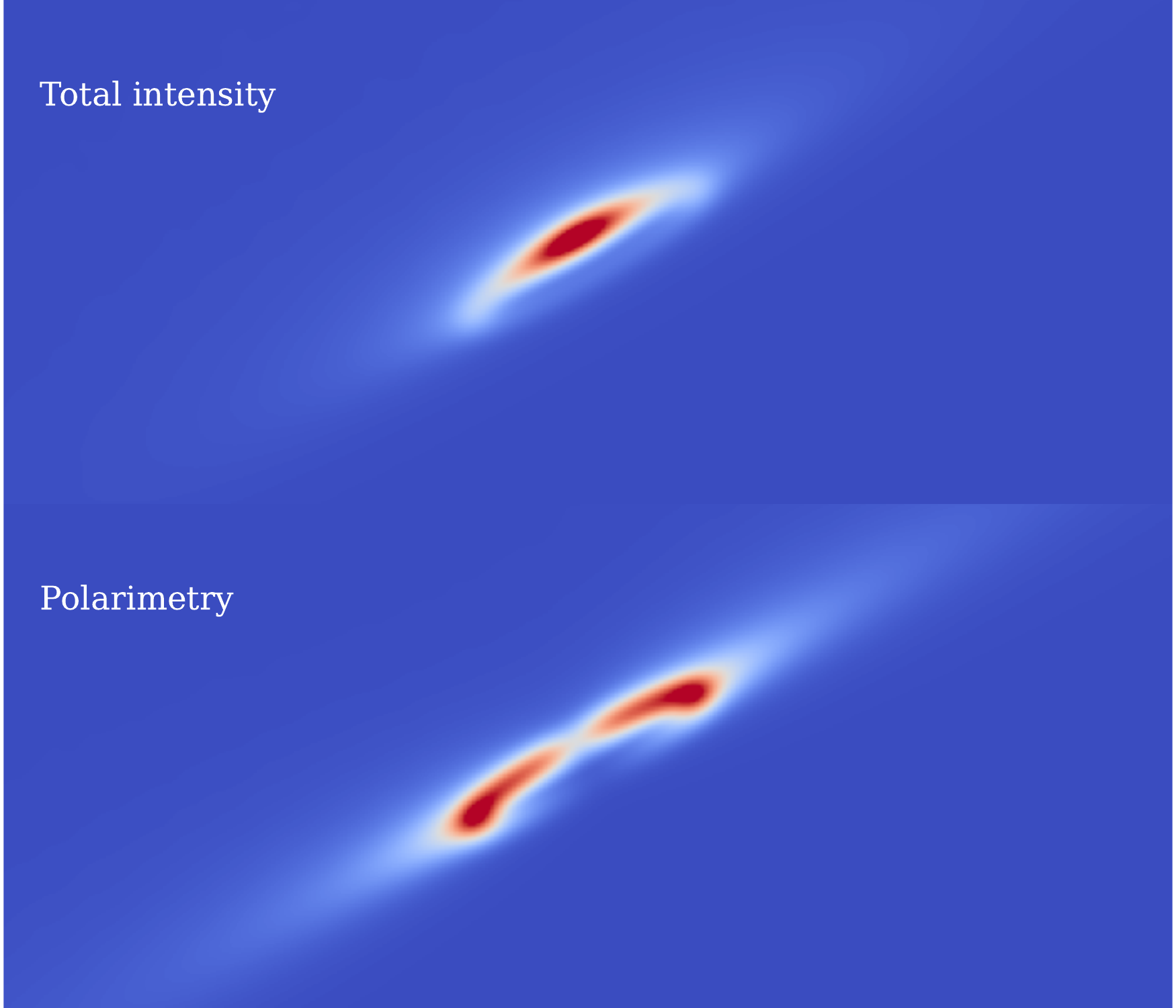}
    \caption{Images in total intensity and polarimetry (top and bottom, respectively), integrated over the full size distribution between $\beta_\mathrm{min}$ and $\beta_\mathrm{max}$, with a square root scaling.
  \label{fig:not_wavy}}
\end{figure}

The origin of this arc, only seen in total intensity, is driven by two main ``parameters''. The first one is a natural consequence of the way the images are computed. By selecting a narrow range of $\beta$ values, we only account for a population of particles that are released from the same distance from the star ($\sim a_0 \pm \delta_\mathrm{a}$, see next paragraph) and have more or less the same eccentricity ($e \sim \beta / (1 - \beta)$). This means that all these particles will have the same apocenter distance (but with arguments of pericenter uniformly distributed between $[0, 2\pi]$). Since the particles will spend most of their time near that location, this results in a pile-up of particles (that we refer to as the ``apocenter pile-up''). If we consider the full size distribution (and not one interval of $\beta$ values), the sum of all these apocenter pile-ups results in a halo with in a surface brightness that naturally follows a power-law (e.g., \citealp{Thebault2008}). Figure\,\ref{fig:not_wavy} shows images integrated over the full size distribution, between $\beta_\mathrm{min}$ and $\beta_\mathrm{max}$, in total and polarized intensity (top and bottom, respectively). The arc-like structure disappears and we retrieve a continuous halo beyond the birth ring (best seen with the square root scaling for both panels). It is also important to keep in mind that in the modeling approach, we assumed that all the grain sizes have scattering efficiencies $Q_\mathrm{sca}$ equal to unity, which is a simplification, that we will further investigate in Section\,\ref{sec:qsca}. Even if the surface density (or optical depth) profile follows a power-law, variations in $Q_\mathrm{sca}$ might indeed affect the resulting appearance of the disk in scattered light observations.

The second ``parameter'' can inform us about the intrinsic properties of the birth ring and is related to the radial width of the birth ring. For the apocenter pile-up to be bright enough to be detected, the range of apocenters for a population of particles with comparable $\beta$ needs to be narrow. The range of the apocenters is positively correlated with the width of the birth ring. Therefore, the detection of an arc strongly suggests a very narrow birth ring where the collisions are taking place (justifying why we used $\delta_\mathrm{a} = 0.025\arcsec$ in Figure\,\ref{fig:adi_dpi}), possibly combined with a variation of the dust properties. But overall, if there is no secondary dust belt, the narrowness of the birth ring is a prerequisite for the detection of an arc.

\subsection{Scattering efficiencies}\label{sec:qsca}

\begin{figure}
  \centering
  \includegraphics[width=\hsize]{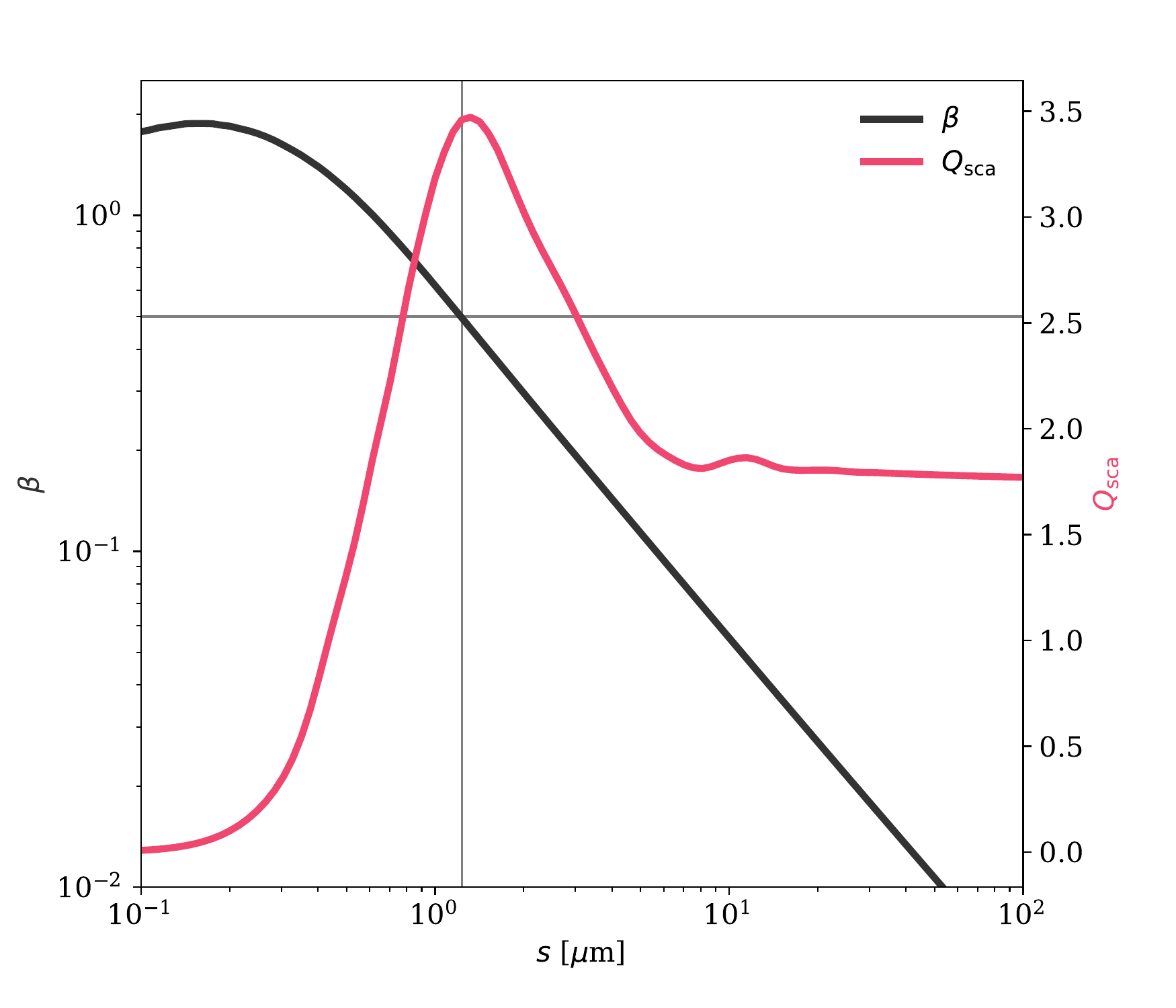}
  \caption{$\beta$ ratio and $Q_\mathrm{sca}$ scattering efficiencies (different $y$ axis) as a function of the grain size. The vertical line denotes the blow-out size for which $\beta = 0.5$ (horizontal line).}
  \label{fig:qsca}
\end{figure}

As mentioned previously, the scattering efficiencies $Q_\mathrm{sca}$ can affect the surface brightness profile of the disk. Particles of some sizes scatter more efficiently the stellar light than others. To have a first order estimate of this effect, we computed both the $\beta$ ratio and the $Q_\mathrm{sca}$ values for a given dust composition, as a function of the grain size, and the results are displayed in Figure\,\ref{fig:qsca}. To compute those values, we used the Distribution of Hollow Spheres model (DHS, \citealp{Min2005}), with a maximum filling factor of $f_\mathrm{max} = 0.8$, and with the ``DIANA'' composition (pyroxene and carbon with a mass ratio of $87$\% and $13$\%, respectively, and a porosity of $25$\%, \citealp{Woitke2016}, optical constants from \citealp{Dorschner1995} and \citealp{Zubko1996}, respectively), using the \texttt{optool}\footnote{Available at \url{https://github.com/cdominik/optool}} package (\citealp{optool}). As explained in \citet{Woitke2016} and references therein, this dust mixture is expected to be well representative of particles in circumstellar disks.

The $\beta$ ratio is computed following \citet{Burns1979}, as
\begin{equation}
  \beta(s) = \frac{3 L_\star}{16 \pi G c^2 M_\star} \times \frac{Q_{\mathrm{pr}}(s)}{\rho s},
\end{equation}
where $M_\star$ and $L_\star$ are the stellar mass and luminosity ($1.3$\,$M_\odot$ and $3.1$\,$L_\odot$, respectively, taken from \citealp{Kral2020}), $G$ the gravitational constant, $c$ the speed of light, and $\rho$ the density of the dust particles. $Q_\mathrm{pr}(s)$ is the radiation pressure efficiency, and is equal to $Q_\mathrm{ext}(s, \lambda) - g_\mathrm{sca}(s) \times Q_\mathrm{sca}(s, \lambda)$ averaged over the stellar spectrum. The $Q_\mathrm{ext}$ and $Q_\mathrm{sca}$ are the extinction and scattering efficiencies and $g_\mathrm{sca}$ the asymmetry parameter, all of which were computed using \texttt{optool}.

While both $\beta$ and $Q_\mathrm{sca}$ depend on the composition and porosity of the particles, for a ``typical'' circumstellar composition (the DIANA standard opacities), Figure\,\ref{fig:qsca} shows an interesting behavior. The blow-out size, $s_\mathrm{blow-out}$ above which the grains remain gravitationally bound to the star, is of about $1.2$\,$\mu$m, and coincidentally this characteristic size also corresponds to a maximum of the scattering efficiency. The $Q_\mathrm{sca}$ values then decrease as $s$ increases, before reaching a plateau for sizes of about $\sim 10$\,$\mu$m. With all other aspects already considered in the code discussed above (size distribution, cross section of the grains, stellar illumination, extended lifetime outside of the birth ring), the contribution to the total flux of the smallest bound particles is enhanced by an additional factor of $\sim 2$ compared to larger particles, solely because of the $Q_\mathrm{sca}$ values.

As discussed at the end of the last section, there are two ``parameters'' that can lead to the detection of an arc-like structure. The first one is to have a population of particles with a narrow range of $\beta$ values. If the size distribution is too wide (see Fig.\,\ref{fig:not_wavy} with the full size distribution) then the arc disappears, but, adding the dependency of $Q_\mathrm{sca}$ with $s$ in the mix might change this result. Even for a wide size distribution, the important $Q_\mathrm{sca}(s)$ variations could indeed increase the contribution to the flux  of a relatively narrow domain of particle sizes close to $s_\mathrm{blow-out}$. There is here, however, a competition between the $Q_\mathrm{sca}$ values and the illumination factor ($\propto 1/r^2$). Indeed, the particles that have large $Q_\mathrm{sca}$ values are also the ones with larger $\beta$ values (as long as particles are bound, that is, $\beta < 0.5$) and are therefore far away from star, receiving less stellar light. As an illustration, we do not see an arc for the particles close to the blow-out size (top right panel of Fig.\,\ref{fig:adi_dpi}) even though they have the largest $Q_\mathrm{sca}$ values (Fig.\,\ref{fig:qsca}). 

\begin{figure}
  \centering
  \includegraphics[width=\linewidth]{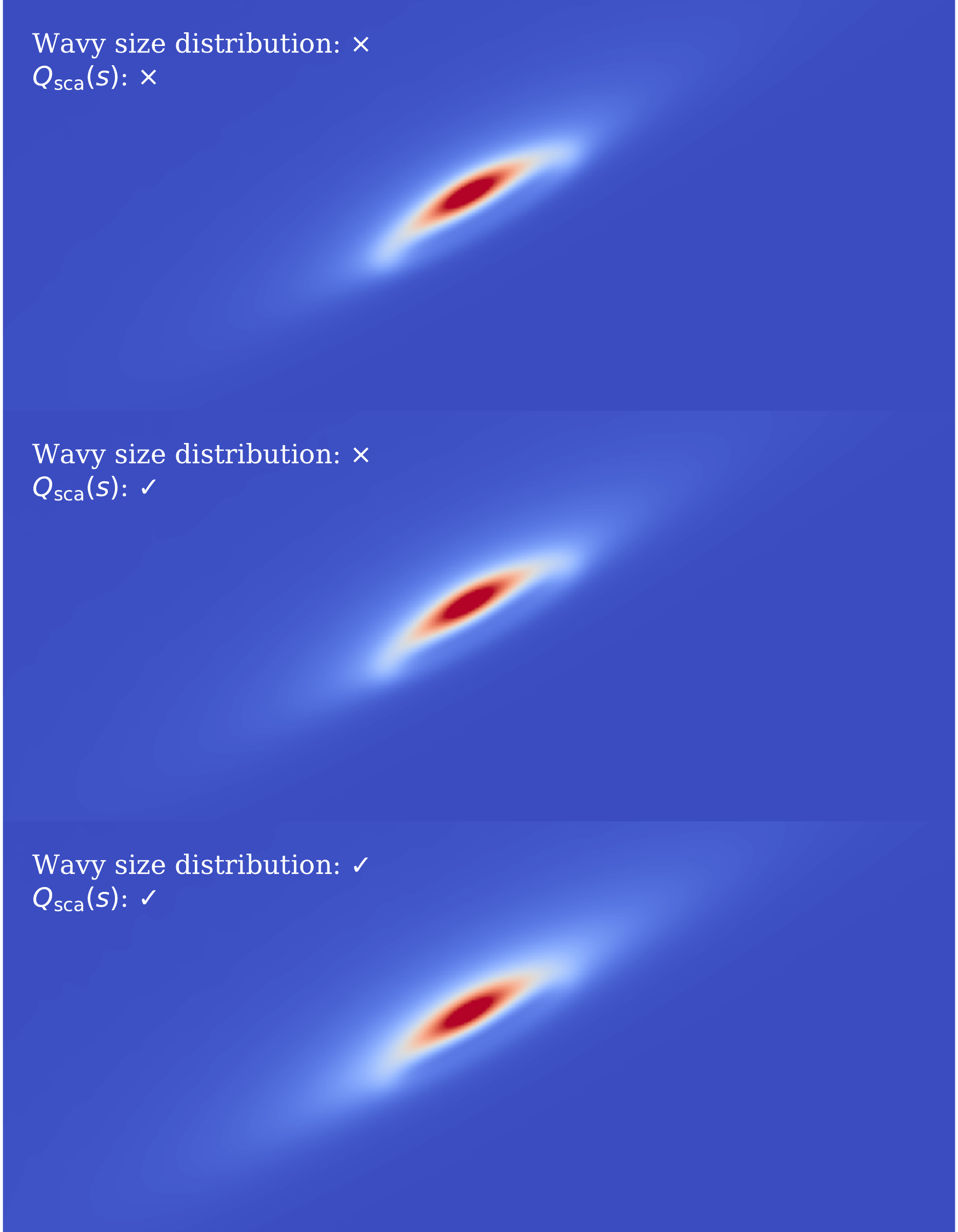}
    \caption{Images in total intensity, integrated over the full size distribution between $\beta_\mathrm{min}$ and $\beta_\mathrm{max}$. The different panels show the differences when accounting for the wavy size distribution or the variation of $Q_\mathrm{sca}$ as a function of $s$. The images are shown with a square root scaling to better highlight the regions outside the birth ring.
  \label{fig:wavy}}
\end{figure}

Figure\,\ref{fig:wavy} shows images in total intensity (see Fig.\,\ref{fig:wavy_dpi} for the same images in polarized intensity). The top panel is the same as for Figure\,\ref{fig:not_wavy}, assuming a size distribution in $\mathrm{d}n(s) \propto s^{-3.5}\mathrm{d}s$ and $Q_\mathrm{sca}$ values equal to unity. In the middle panel, we take into account the variation of $Q_\mathrm{sca}$ as a function of the grain size $s$ (Fig.\,\ref{fig:qsca}), and the bottom panel is further discussed in Section\,\ref{sec:wavy}. The images have a square root scaling to better highlight the faint halo and overall the differences between the top and middle images are marginal. The variation in $Q_\mathrm{sca}$ as a function of $s$ does not seem to be sufficient to solely explain the presence of an arc.

\subsection{A size-dependent phase function}\label{sec:size_pfunc}

\begin{figure}
  \centering
  \includegraphics[width=\hsize]{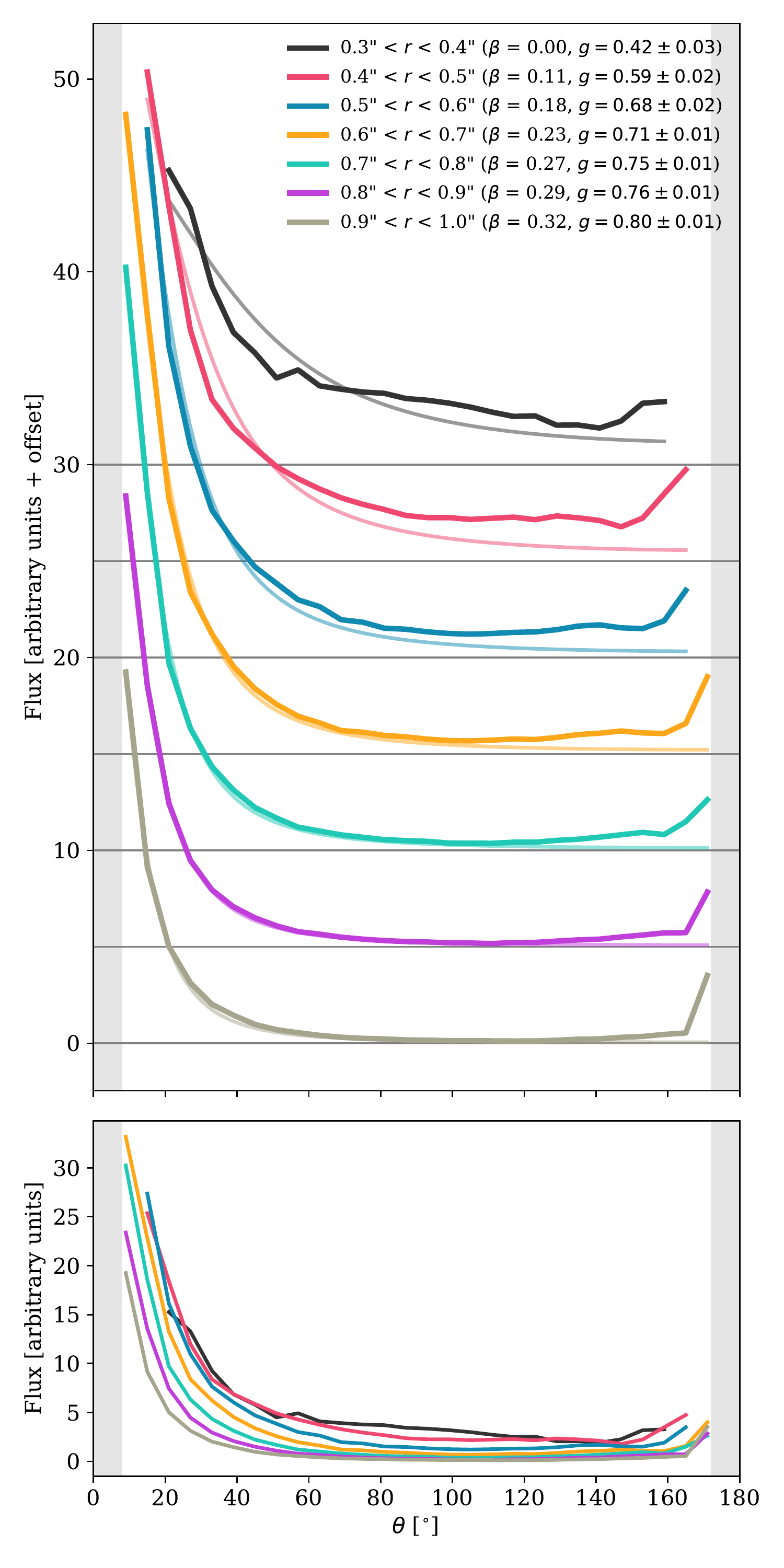}
    \caption{Surface brightness as a function of the scattering angle, extracted in concentric annulii (accounting for the inclination and position angle of the disk) with increasing deprojected stellocentric distances $r$ (in arcsec). For clarity, the profiles have been offset and the horizontal lines show the zero point on the top panel, while no offset has been included for the bottom panel to best compare the different profiles. The fainter lines on the top panel show the best fit with an HG phase function (the $g$ values being reported in the legend). The shaded gray areas show the scattering angle that are not accessible to us for an inclination of $82^\circ$.}
  \label{fig:pfuncr}
\end{figure}

Besides these considerations, there is also a geometrical issue, which is that the arc seen in our simulated images appears slightly more ``curved'' compared to the one seen in the observations (e.g., third or fourth panels of Fig.\,\ref{fig:adi_dpi}). This is because, assuming the birth ring is circular (\citealp{Olofsson2022} could reproduce the $Q_\phi$ observations without needing an eccentric disk), the apocenter pile-up for a narrow range of $\beta$ will also be circular. For each panel of Fig.\,\ref{fig:adi_dpi} we are only considering narrow intervals of $\beta$ values, and the same phase function for all of them ($g = 0.7$). But the phase function does depend on the grain size (and as a consequence of radiation pressure, on the stellar distance). In total intensity, grains much smaller than the wavelength should in principle display isotropic scattering, while grains larger or comparable to the wavelength should scatter in the forward direction more efficiently. Coupled with the efficient size segregation caused by radiation pressure, the integrated phase function at different distances from the star should vary, which could in the end alter the shape of the arc.

There is in fact a way to check if our observations point towards a phase function that varies with stellar distance. This can be done by computing, at different radial distances, the brightness profile as a function of the scattering angle in the midplane (assuming the disk is vertically thin\footnote{See the discussion in \citealp{Olofsson2020} on the impact of the vertical height on the phase function. In the case of HD\,129590, \citet{Kral2020} reported the presence of gas and \citet{Olofsson2022b} showed that gas-bearing disks are most likely vertically thin at near-IR wavelengths, justifying the assumption of a thin disk.}). This is done in concentric annuli accounting for the inclination and position angle of the disk (hence deprojected stellocentric distances). We binned the scattering angles in $30$ bins between $0$ and $180^\circ$, and for each bin we selected the pixels that have a midplane scattering angle within the range of the considered bin. We then computed the mean value of the selected pixels. Figure\,\ref{fig:pfuncr} shows the extracted profiles for $7$ annuli, with a width of $0.1\arcsec$ between $0.3\arcsec$ and $1\arcsec$ (deprojected distances). Since we masked the innermost regions of the image ($0.125\arcsec$), the profiles for the first three annuli cover a smaller range of scattering angles. On the top panel, we added an offset to the profiles for clarity, and the horizontal lines show the zero point, while the bottom panel shows the profiles without the offset for a direct comparison between them. Furthermore, for each annulus, we used \texttt{lmfit} (\citealp{lmfit}) to find the HG phase function (Eqn.\,\ref{eqn:pfunc}) with the $g$ value that best fits the profile. The best fitting values are reported in the legend of Fig.\,\ref{fig:pfuncr} and the best HG fits are shown with dimmer lines underneath the extracted profiles on the top panel. In the legend, we also provide an estimate of the $\beta$ values of the particles that dominate the geometrical cross section for each range of distance. To compute the $\beta$ values, we assume that between the deprojected distances $r_1$ and $r_2$ (e.g., $0.7$ and $0.8\arcsec$), we are seeing particles that have their apocenter at this distance from the star, while their pericenter lies in the birth ring ($r_0 = 0.35\arcsec$, \citealp{Olofsson2022}). We can compute $\bar{r} = (r_1 + r_2)/2$, and derive the eccentricity $e = (\bar{r} - r_0)/(\bar{r} + r_0)$, from which we derive the ``typical'' $\beta$ value as $e/(1+e)$. This is only done to provide a rough estimate of the typical grain size but the $\beta$ values are not used when extracting the profiles. 

For all $7$ profiles, we can first notice on Figure\,\ref{fig:pfuncr} that there may be some back-scattering that the Henyey-Greenstein functions fails to capture (since we used only one value for $g$), for scattering angles larger than $\sim 150^{\circ}$. This back-scattering could be real, but could also be due to an improper subtraction of the reference images, leaving some residual flux near the center of the image. It should also be noted that the HG fits to the first two brightness profiles ($[0.3,0.4\arcsec]$ and $[0.4, 0.5\arcsec]$) do not reproduce very well their shape but that the fits are quite decent for the other ones (beyond $\sim 0.5-0.6\arcsec$). Interestingly, we can see that the value of $g$ is increasing the farther away we go from the birth ring, and this result is quite counter-intuitive. We should expect small particles (large $\beta$) to scatter isotropically, leading to lower values of $g$, while we see the opposite trend, with $g$ reaching a value of $0.8 \pm 0.01$ for the largest $\beta$ values. The difference with the values of $g$ reported in \citet[][$g = 0.42-0.52$]{Matthews2017} can be explained by the better recovery of the semi-minor axis of the disk in the DI-sNMF reduction compared to the PCA reduction.

While this seems counter-intuitive at first, things are not that straightforward. We must keep in mind that our assumption of a single grain size at a given distance in the halo is not fully true. At a given distance outside the birth ring, we still probe a distribution of sizes. For instance, the smallest bound grains contribute to the flux of all the concentric annuli, not only in the outermost annulus where their apocenter is. On top of that, for an inclination of $i = 82^\circ$ we are probing scattering angles in the range $[90^\circ - i, 90^\circ+i]$, meaning that we are missing the very small (and large) scattering angles (the missing angles are marked by the shaded gray areas in Fig.\,\ref{fig:pfuncr}). For grains as large as $\sim 5$\,$\mu$m, the strong forward scattering peak can fall in this ``blind'' range (see \citealp{Mulders2013}). Therefore, by missing the forward scattering peak, the phase function of large grains might in fact appear almost isotropic in the range of angles we are probing\footnote{As discussed in \citet{Mulders2013}, this should also impact the $Q_\mathrm{sca}$ values, but for the sake of simplicity we opted not to include this effect in our calculations.}. Our results seem to indicate that the smallest dust grains that our observations are sensitive to, have a phase function with $g \sim 0.8$ (bottom profile in Fig.\,\ref{fig:pfuncr}, between $0.9$ and $1$$\arcsec$). As we get closer and closer to the birth ring, larger grains are also contributing to the flux, and these particles might have an isotropic ``effective'' phase function as we are missing the strong forward scattering peak. This results in an integrated phase function with lower $g$ values, as the contribution of the larger grains keeps on effectively adding flux at all scattering angles.

Overall, this discussion remains to be further tested. There are some aspects that our procedure does not take into account, such as the vertical scale height (\citealp{Olofsson2020}), the effects of the convolution and of the inclination of the disk when extracting the profiles (elliptic apertures, \citealp{Milli2017}), or using more realistic phase functions than the HG approximation (Eqn.\,\ref{eqn:pfunc}). Nonetheless, variations of the phase function as a function of the size, hence $\beta$, hence radial distance, is something that could possibly explain the shape of the arc as seen in the observations. Indeed, as can be seen in Fig.\,\ref{fig:pfuncr}, for scattering angles in the range $25-50^{\circ}$ the annuli become relatively brighter as $r$ decreases (bottom to top). This widens the wings along the minor axis of the disk, thus altering the shape of the arc. Furthermore, as the quality of the observations and post-processing algorithms keep on improving, using a single phase function for the whole disk (birth ring and halo) might become a limitation when modeling the scattered or polarized light observations. The change in the size distribution as we venture away from the birth ring might bias our characterization of the phase functions of young debris disks.

\subsection{A wavy size distribution}\label{sec:wavy}

For the apocenter pile-up to produce a detectable arc, we need to have a rather sharp variation for one of the following two variables. We either need a strong peak in the size distribution, to ``isolate'' a particular particle size, or an equally strong peak in the scattering efficiencies (resulting in a similar ``isolation'' of a given grain size, by making all the other sizes dimmer). Regarding the latter explanation it is challenging to think that there could be a drop in $Q_\mathrm{sca}$ sharper than the one displayed in Fig.\,\ref{fig:qsca}.

A result that might help explaining the apocenter pile-up was presented in \citet{Thebault2007}, who demonstrated that the size distribution can display some wavy structures. The origin of this waviness comes from the fact that grains with a size $s$ can efficiently destroy particles with a size $s + \delta s$. Because of the radiation pressure there is a lack of grains with sizes $s \lesssim s_\mathrm{blow-out}$, leading to an over-abundance of particles with $s \gtrsim s_\mathrm{blow-out}$ which can efficiently destroy slightly larger grains. As time passes, this pattern of over/under-abundance propagates throughout the whole size distribution. However, the authors showed that the resulting pattern is rather smooth, and the first ``peak'' can extend in the range $\sim 5-10$\,$\mu$m in their simulations. But, interestingly, their resulting mass distribution shows a very steep increase right before the first peak, which is not located at $s_\mathrm{blow-out}$ but at slightly larger sizes ($s_\mathrm{peak} \sim 1.5 \times s_\mathrm{blow-out}$). Assuming $\beta \propto 1/s$ and that $\beta_\mathrm{blow-out} = 0.5$, we find that $\beta_\mathrm{peak} \sim 0.33$. Going back to Figure\,\ref{fig:adi_dpi}, the first peak of the wavy size distribution would then correspond to the second to last upper panel (with $\beta \sim 0.35$). The contribution of grains with larger $\beta$ values would be severely diminished (rightmost upper panel) because of the wavy distribution, the ideal configuration for the arc to naturally appear in the observations. Particles with smaller $\beta$ can still remain in the system, but their apocenter will never reach as far out, and they therefore would not be able to dilute the apocenter pile-up. In other words, this could work if the size distribution is effectively cut off at a slightly larger size than the blow-out size, possibly combined with larger $Q_\mathrm{sca}$ values for smaller particles.

The bottom panel of Figure\,\ref{fig:wavy} shows the image in total intensity when also accounting for the wavy size distribution (following Eqn.\,5 from \citealp{Thebault2007}). The main difference between the three images is that the halo seems to be slightly brighter when including the wavy size distribution along with the scattering efficiencies, and seems to be slightly offset compared to the projected major axis of the disk. However, the square root scaling is necessary to reveal the halo, which remains faint. This could be due to the choice of the dust composition and the variation of $Q_\mathrm{sca}$ that is not sufficient to make the halo brighter, or this could be due to the shape of the wavy size distribution. In Fig.\,\ref{fig:qsca} the peak of $Q_\mathrm{sca}$ corresponds to $s \sim s_\mathrm{blow-out}$, but if a different composition were to shift the peak toward slightly larger sizes (e.g., close to $\beta \sim 0.35$), this might help increasing the contribution of the arc. Furthermore, Equation\,5 from \citet{Thebault2007} is meant for radially broader disks and might not be directly applicable for narrower birth rings. To conclude on this exercise, it goes in the right direction even though the magnitude of the effect might not be sufficient to reproduce the observations. Another additional effect that might help strengthening the signal from the arc is to also include the variation of the phase function with the grain size. Unfortunately, this cannot be easily tested since we can hardly quantify this variation. Figure\,\ref{fig:pfuncr} might seem to be a good starting point, but we have to keep in mind that the derived values of $g$ are not for a single grain size. As mentioned before, for each concentric annuli there is a distribution of grain sizes contributing to the surface brightness and we cannot easily determine the relationship between $g$ and $\beta$ (to determine $g(s)$).

Overall, explaining why we would detect only a narrow range of particle sizes (either because of the underlying size distribution or because of the optical properties) remains quite a significant challenge. But the fact remains that the total and polarized intensity observations of HD\,129590 show different structures. Furthermore, the fact that the arc is actually seen in the reduction of \citet[][different team, dataset, reduction process]{Matthews2017} brings confidence that it is indeed real and needs to be explained.

\subsection{The width of the birth ring}

As mentioned in Section\,\ref{sec:pfunc}, for the apocenter pile-up to produce an arc, the birth ring must be narrow enough so that the apocenters of particles of comparable sizes are not too radially diluted. It can therefore be an indirect way to hint that the parent planetesimals are indeed arranged in a ring with a small range of semi-major axis. A clear follow-up observation would be to image HD\,129590 with ALMA at high angular resolution (the beam size in the observations presented in \citealp{Kral2020} was $1.61 \times 1.16\arcsec$) and assess whether the birth ring is indeed narrow or radially extended.


\section{Perspectives and discussion}

\subsection{Total intensity versus polarimetry}

The results presented in this paper add to the already known differences between total and polarized intensity observations of debris disks. Simply because of the different shapes of the phase function, total intensity observations usually best trace the projected minor axis of the disks, while their major axis is usually best revealed in polarimetric observations. But as demonstrated above, the differences between total and polarized intensity can be even more profound and can in fact strongly bias our interpretation of the structure of the disk. A disk can appear as a single belt in polarimetry, but a faint, additional arc might appear in total intensity. We showed that this arc should not always be interpreted as the presence of a secondary belt, as it can instead be a mirage caused by the combination of radiation pressure and a more forward scattering phase function.

As a side note, the presence of the arc in the total intensity image might slightly bias the determination of the inclination of the disk. \citet{Matthews2017} determined an inclination of $\sim 75^{\circ}$ from their total intensity reduction, while \citet{Olofsson2022} obtained an inclination of $i \sim 82^{\circ}$ when modeling the polarized intensity observations. In total intensity, as the arc shifts the photocenter along the direction of the minor axis of the disk it can make the disk appear less inclined than it really is. 

For these reasons, polarized observations allow us to more reliably constrain the actual morphology and geometry of debris disks (on top of being less affected by post-processing algorithms).

\subsection{Arcs in other debris disks}\label{sec:otherdisks}

If our interpretation for the detection of the arc-like structure holds for HD\,129590, one may wonder why such arcs are not routinely detected in other debris disks. As mentioned previously, there is at least one prerequisite; the birth ring must be quite narrow so that the pile-up at the apocenter is not too spread out. However, as mentioned in \citet{Marino2022}, results from the REASONS survey (Matr\`a et al. in prep) suggest that debris disks are in general quite extended radially, with a median fractional width ($\Delta r/r$) of about $0.74$, much wider than the Kuiper belt of the solar system for instance ($\sim 0.27$). This would suggest that the disk around HD\,129590 is to some extent different than most of the other debris disks. This could explain why such structures are not routinely detected.

But the case of HD\,120326 (HIP\,67497) is an interesting comparison point. \citet{Bonnefoy2017} reported the detection of an inclined disk ($i \sim 80^{\circ}$) as well as an arc in the SPHERE total intensity observations, and investigated several scenarios. They first modeled the inner belt only (corresponding to the birth ring in our case), assuming steep power-law distributions for the volumetric density, $\alpha_\mathrm{in} = 10$ and $\alpha_\mathrm{out} = -5$, a rather narrow birth ring located at $58.6$\,au. They then tested two models, to be added to this inner belt, an extended halo or another birth ring (with a reference radius different than the one of the first belt). Based on the residuals they concluded that the latter scenario was more likely, and for this secondary belt, they found a radius of $130$\,au and an outer slope in $\alpha_\mathrm{out} = -8$, a sharp outer edge for the secondary ring. From their images, the secondary ring appears quite farther from the innermost one, and a ``gap'' between the two can be noticed. Nonetheless, \citet{Olofsson2022} presented polarimetric observations of the same disk, and no secondary ring is detected in the observations, suggesting that HD\,120326 could well be a twin (or maybe a relative) of HD\,129590.

A possible counter-example, a narrow and inclined birth ring but with no apparent secondary ring or arc, is HR\,4796. The disk has been intensively monitored at near-IR wavelengths, both in total and polarized intensity, and no arc \textit{in the vicinity} of the main ring has ever been reported. But \citet{Schneider2018} presented the detection of ``extensive exo-ring dust material'' from Hubble Space Telescope observations, and that this exo-ring is located much farther away from the birth ring.

For the sake of the argument, let us assume that for these three objects (HD\,129590, HD\,120326, and HR\,4796) the arcs can be solely explained by the apocenter pile-up. One of the main difference between these three stars is that their spectral types are G3, F0, and A0, respectively. In Section\,\ref{sec:wavy}, we discussed the possible impact of a wavy size distribution as described in \citet{Thebault2007}. However, the authors investigated the effect that the central stellar mass could have on their results and found the differences to be negligible. Their Figure\,7 shows that the size distribution is simply shifted depending on the value of $s_\mathrm{blow-out}$, but the shape of the peak or of the steep increase before seem to remain similar. It would therefore be challenging to explain why the arc around HR\,4796 is located so far away from the birth ring.

\section{Summary}

In this paper, we reported on the detection of an arc-like structure in the disk around HD\,129590. This arc is only detected in total intensity and not in polarimetric data. After discussing why it is unlikely related to a secondary ring, we proposed a possible mechanism that could be responsible for its detection. A promising candidate for such a mechanism would be the apocenter pile-up, in which we are detecting the apocenter of particles of a given size. While this remains challenging to explain, this could be due to some truncation of the size distribution at $\beta$ values smaller than the expected cut-off size of $\beta \sim 0.5$ (otherwise the apocenters are too far away for the pile-up to be detected). We hypothesize this could be the consequence of a wavy size distribution, possibly combined with the dependence of $Q_\mathrm{abs}$ as a function of the particle size $s$.

A requirement for the apocenter pile-up to be detected is that the birth ring needs to be radially narrow so that the apocenters are not too diluted. Given that most disks have rather broad birth rings, this could explain why such structures are not routinely detected, but the disks around HD\,129590 and HD\.120326 (and possibly HR\,4796) are interesting candidates to further investigate this effect. Future high angular resolution ALMA observations would help to confirm the width of the planetesimals belt of the disk around HD\,129590.

\begin{acknowledgements}
We thank the anonymous referee for a constructive and helpful report, that helped strengthen the results of this paper. J.\,O., A.\,B., B.\,M.-O., and N.\,G. acknowledge support by ANID, -- Millennium Science Initiative Program -- NCN19\_171.
K.\,M. is funded by the European Union (ERC, WANDA, 101039452). Views and opinions expressed are however those of the author(s) only and do not necessarily reflect those of the European Union or the European Research Council Executive Agency. Neither the European Union nor the granting authority can be held responsible for them.
SPHERE is an instrument designed and built by a consortium consisting of IPAG (Grenoble, France), MPIA (Heidelberg, Germany), LAM (Marseille, France), LESIA (Paris, France), Laboratoire Lagrange (Nice, France), INAF–Osservatorio di Padova (Italy), Observatoire de Gen\`eve (Switzerland), ETH Zurich (Switzerland), NOVA (Netherlands), ONERA (France) and ASTRON (Netherlands) in collaboration with ESO. SPHERE was funded by ESO, with additional contributions from CNRS (France), MPIA (Germany), INAF (Italy), FINES (Switzerland) and NOVA (Netherlands).  SPHERE also received funding from the European Commission Sixth and Seventh Framework Programmes as part of the Optical Infrared Coordination Network for Astronomy (OPTICON) under grant number RII3-Ct-2004-001566 for FP6 (2004–2008), grant number 226604 for FP7 (2009–2012) and grant number 312430 for FP7 (2013–2016). We also acknowledge financial support from the Programme National de Plan\'etologie (PNP) and the Programme National de Physique Stellaire (PNPS) of CNRS-INSU in France. This work has also been supported by a grant from the French Labex OSUG@2020 (Investissements d'avenir – ANR10 LABX56). The project is supported by CNRS, by the Agence Nationale de la Recherche (ANR-14-CE33-0018). It has also been carried out within the frame of the National Centre for Competence in Research PlanetS supported by the Swiss National Science Foundation (SNSF). MRM, HMS, and SD are pleased to acknowledge this financial support of the SNSF. 
This work has made use of data from the European Space Agency (ESA) mission {\it Gaia} (\url{https://www.cosmos.esa.int/gaia}), processed by the {\it Gaia} Data Processing and Analysis Consortium (DPAC, \url{https://www.cosmos.esa.int/web/gaia/dpac/consortium}). Funding for the DPAC has been provided by national institutions, in particular the institutions participating in the {\it Gaia} Multilateral Agreement.
This research made use of Astropy,\footnote{\url{http://www.astropy.org}} a community-developed core Python package for Astronomy \citep{astropy:2013, astropy:2018}, Numpy (\citealp{numpy}), Matplotlib (\citealp{matplotlib}), Scipy (\citealp{scipy}), and Numba (\citealp{numba}).
\end{acknowledgements}

\bibliographystyle{aa}

\appendix

\section{Alternative data reduction}

\begin{figure*}
  \centering
  \includegraphics[width=\linewidth]{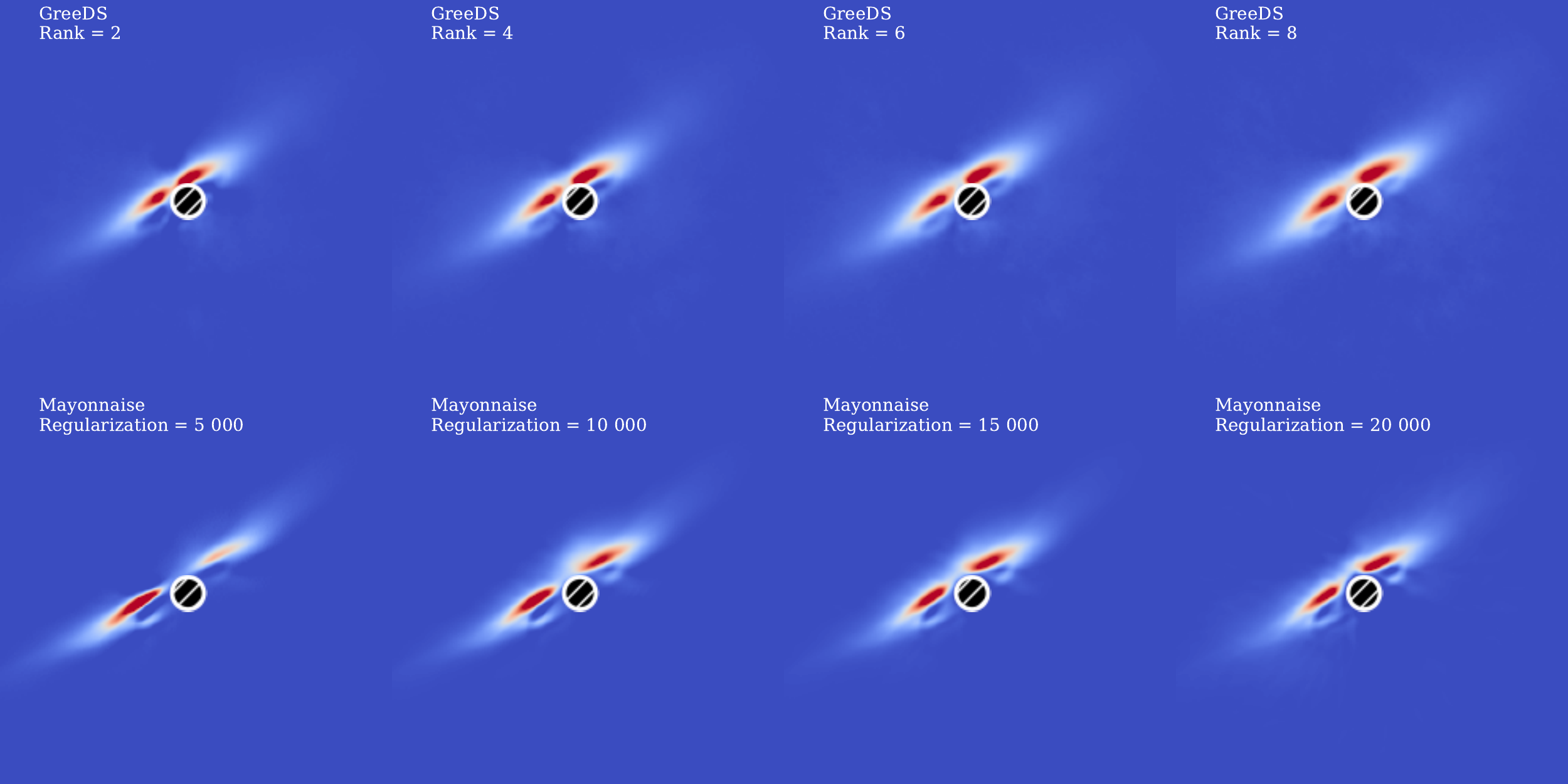}
    \caption{Data processing using the \texttt{GreeDS} and \texttt{MAYONNAISE} algorithms (top and bottom, respectively). The main parameters for the reduction (rank and regularization parameters) are reported in each image. The scaling is linear up to the $99.9$ percentil, and re-evaluated for each panel.
  \label{fig:mayo}}
\end{figure*}

The top and bottom rows of Figure\,\ref{fig:mayo} respectively show the reduction of the science data cube using the \texttt{GreeDS} and \texttt{MAYONNAISE} algorithms. The interested reader is referred to the original papers describing the pipelines, \citet{Pairet2018} and \citet{Pairet2021}. The main parameters for the different reduction are reported in the individual images of the Figure\,\ref{fig:mayo}, and are the rank number for \texttt{GreeDS} and the disk regularization for \texttt{MAYONNAISE} (using a rank $6$ for the \texttt{GreeDS} preliminary step). The other parameters are set to the defaults values, a central mask of $8$\,pixels, $10\,000$ iterations, and a tolerance threshold of $10^{-4}$ for the convergence of the algorithm. The images do not recover the flux of the disk along the minor axis as well as the DI-sNMF reduction (Fig.\,\ref{fig:data}), but the arc-like structure close to the major axis of the disk is recovered.

\section{Additional images}

\begin{figure*}
  \centering
  \includegraphics[width=\linewidth]{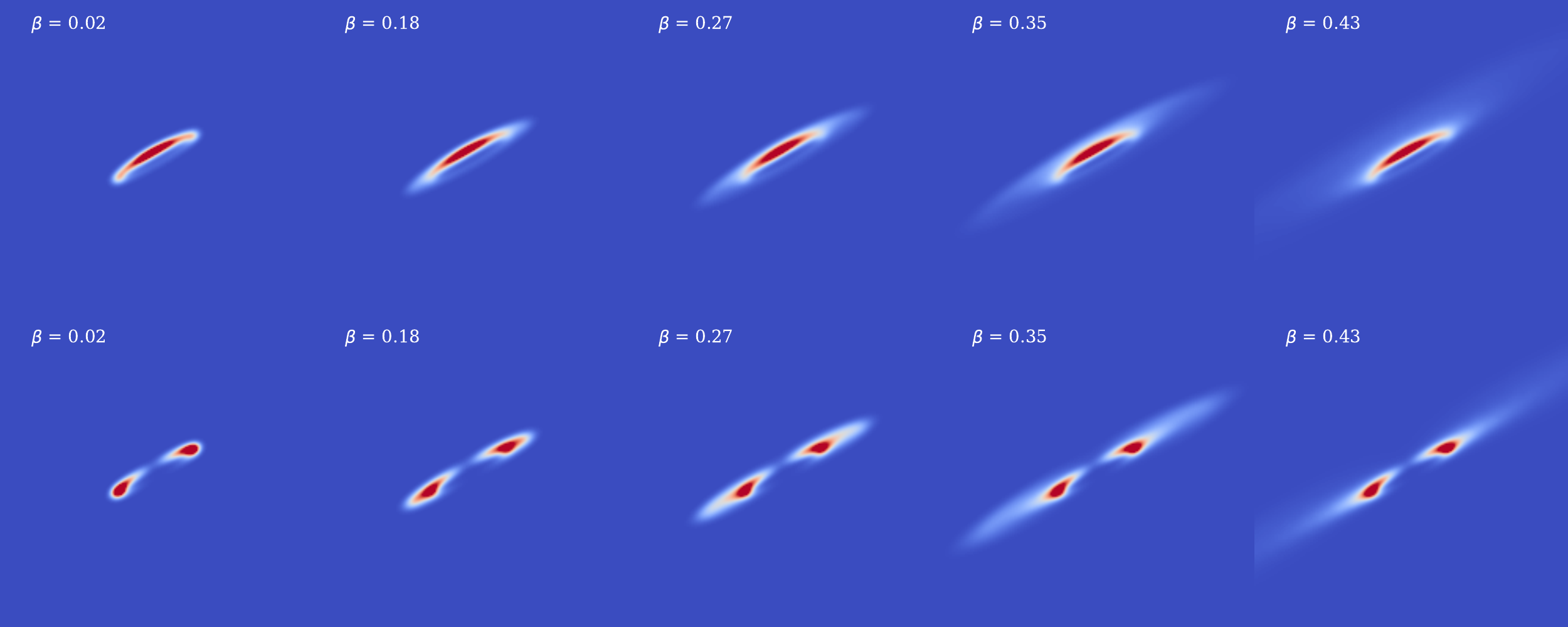}
    \caption{Same as Figure\,\ref{fig:adi_dpi}, but for $g=0.5$ instead of $0.7$.
  \label{fig:adi_dpi05}}
\end{figure*}

Figure\,\ref{fig:adi_dpi05} shows simulated images in total intensity and polarimetry using an HG phase function with $g = 0.5$ (similar to the results from \citealp{Matthews2017}), showing that the arc survives a change in the asymmetry parameter.

\begin{figure}
  \centering
  \includegraphics[width=\linewidth]{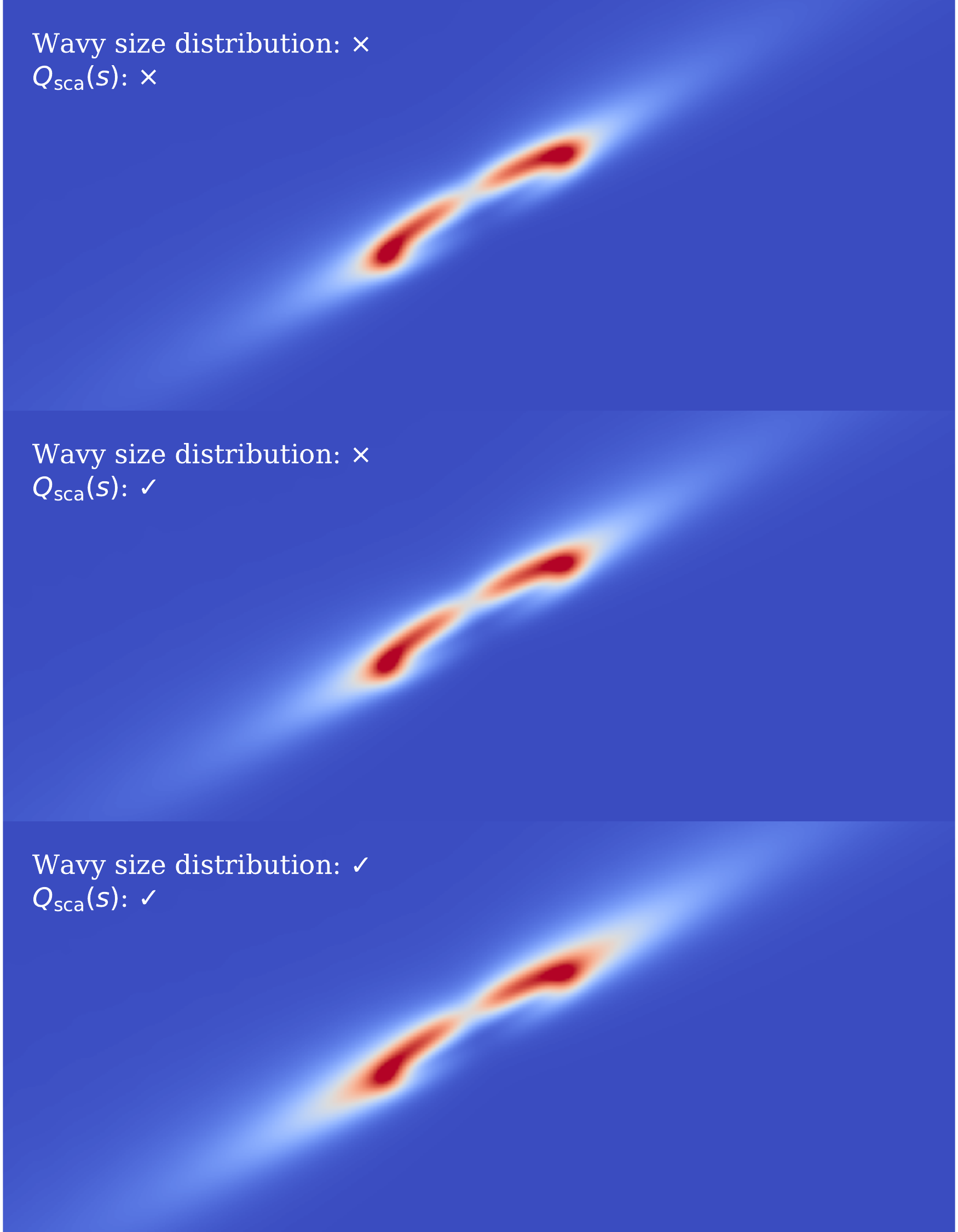}
    \caption{Same as Figure\,\ref{fig:wavy}, but in polarized intensity.
  \label{fig:wavy_dpi}}
\end{figure}

Figure\,\ref{fig:wavy_dpi} shows images in polarized intensity, integrated over the size distribution, when taking into account the variation of $Q_\mathrm{sca}$ with the grain size, and the waviness of the size distribution. It should be noted that the simulated images that include both the effects of $Q_\mathrm{sca}$ and the wavy size distribution look more similar to the polarimetric observations than the images in the middle and top panels.

%

\end{document}